\documentclass[11pt]{article}
\usepackage{epsfig}
\usepackage{graphicx}
\usepackage{amssymb}      
\usepackage{here}  

\evensidemargin=\oddsidemargin  

\newcommand{\beq}{\begin{equation}}
\newcommand{\eeq}{\end{equation}}
\newcommand{\beqa}{\begin{eqnarray}}
\newcommand{\eeqa}{\end{eqnarray}}

\def\O{{\cal O}}
\def\TT{{\cal T}}

\def\End{\end{document}}
\def\thisday{October, 2001}

\def\to{\rightarrow}

\def\dis{\displaystyle}
\def\f{\frac}

\def\[{\left[}
\def\]{\right]}
\def\({\left(}
\def\){\right)}

\def\Ah{\widehat{A}}
\def\cut{\Lambda}
\def\l{{\ell}}

\def\ct{\cos\theta}

\begin{document}

\begin{titlepage}
\def\thepage {}        % Kill page numbering

\title{
\vspace*{7mm}
{\bf Unitarity of Compactified Five Dimensional}\\ 
{\bf Yang-Mills Theory}
\\[7mm]  }

\author{
{\sc R. Sekhar Chivukula}\,$^1$,~~~
{\sc Duane A. Dicus}\,$^2$,~~~
{\sc Hong-Jian He}\,$^2$\,\thanks{Electronic addresses:
     sekhar@bu.edu,
     phbd057@utxvms.cc.utexas.edu,
     hjhe@physics.utexas.edu}\\[5mm]
$^1$\,Department of Physics, Boston University, \\
590 Commonwealth Ave., Boston, Massachusetts  02215 \\[2mm]
$^2$\,Center for Particle Physics,\\ 
University of Texas at Austin, Texas 78712}

\date{\thisday}

\maketitle

\bigskip
\begin{picture}(0,0)(0,0)
\put(360,295){BUHEP-01-27}
\put(360,280){UTHEP-01-25}
\put(360,265){hep-ph/0111016}
%\put(360,280){hep-ph/0111016}
\end{picture}
%\vspace{-12pt}
\vspace{24pt}

\begin{abstract}
  
  Compactified five dimensional Yang-Mills theory results in an effective
  four-dimensional theory with a Kaluza-Klein (KK) tower of massive vector
  bosons.  We explicitly demonstrate that the scattering of the massive
  vector bosons is unitary at tree-level for low energies, and analyze the
  relationship between the unitarity violation scale in the KK theory and the
  nonrenormalizability scale in the five dimensional gauge theory.  In the
  compactified theory, low-energy unitarity is ensured through an interlacing
  cancellation among contributions from the relevant KK levels.  Such
  cancellations can be understood using a Kaluza-Klein equivalence theorem
  which results from the geometric ``Higgs'' mechanism of compactification.
  In these theories, the unitarity violation is delayed to energy scales
  higher than the customary limit through the introduction of additional
  vector bosons rather than Higgs scalars.

\pagestyle{empty}
\end{abstract}
\end{titlepage}

%%%%%%%%%%%%%%%%%%%%%%%%%%%%%%%%
\setcounter{section}{0}
%--\section{Introduction}
%--\setcounter{equation}{0}
%\bigskip

The visible four-dimensional world may be only part of a higher
dimensional space-time structure, with the extra spatial dimensions
substantially larger than the traditional Planck length ($10^{-33}$\,cm),
but too small to have been probed experimentally 
\cite{Antoniadis:1990ew,Antoniadis:1993fh,Lykken:1996fj,Antoniadis:1997hk,
Arkani-Hamed:1998rs,Antoniadis:1998ig,Shiu:1998pa,Dienes:1998vh,Dienes:1998vg,
Pomarol:1998sd,Cheng:1999fu,Lykken:1999ms}.
If the gauge particles propagate in the higher-dimensional space, then from
the four-dimensional viewpoint each gauge boson is associated with a
Kaluza-Klein (KK) tower of massive vector bosons whose mass splittings are
characterized by the (inverse) size of the extra dimensions. In this way
compactification of higher dimensions leads to a 
``geometrical'' mechanism for producing massive vector states.

The high-energy behavior of the scattering of longitudinally-polarized
massive vector bosons is potentially problematic in Yang-Mills theories, 
and can result in amplitudes growing with energy at tree-level
\cite{LlewellynSmith:1973ey,Dicus:1973vj,Cornwall:1973tb,Cornwall:1974km}.  
In the four-dimensional (4D) standard model these amplitudes are exactly 
canceled by the exchange of spin-0 Higgs particle
\cite{Dicus:1973vj,Lee:1977yc,Lee:1977eg,Veltman:1977rt}.
However, such Higgs scalar states do not exist in a 
compactified pure gauge theory.

In this letter, we discuss the high-energy behavior of massive vector-boson
scattering in the compactified five-dimensional (5D) Yang-Mills theory.  We
explicitly demonstrate that the scattering of the massive vector bosons is
unitary at tree-level for low energies, and analyze the relationship between
the scales of 4D unitarity violation and the nonrenormalizability of the 5D
gauge theory.  In the compactified theory we show that the low-energy
unitarity is ensured through an interlacing cancellation among contributions
from the relevant KK levels. We observe that this cancellation can be
understood from a Kaluza-Klein equivalence theorem resulting from the
geometric ``Higgs'' mechanism of compactification.  As a consequence, the
unitarity violation is delayed to energy scales higher than the customary
limit of Dicus-Mathur and Lee-Quigg-Thacker
\cite{Dicus:1973vj,Lee:1977yc,Lee:1977eg,Veltman:1977rt} through the
introduction of additional vector bosons rather than Higgs scalars.

The Lagrangian for five-dimensional Yang-Mills theory is given by
\beq
{\cal L}_5 
= -{1\over 2} {\rm Tr}(\widehat{F}_{MN} \widehat{F}^{MN})\,,~~ \ \ \ 
\widehat{F}^a_{MN} 
= \partial_M \Ah^a_N - \partial_N \Ah^a_M + g^{~}_5
C^{abc} \Ah^b_M \Ah^c_N \,,
\label{fivedym}
\eeq
where $a$ is the gauge index, $C^{abc}$ 
the structure constant, and $g^{~}_5$ the
5D gauge coupling with dimension of (mass)$^{-1/2}$. The
five-dimensional coordinates are labeled by $M,N\in (\mu,\,5)$
with $\mu \in (0,1,2,3)$.

For convenience, we may consider this 5D theory 
with a covariant gauge-fixing term\,\cite{Dienes:1998vg},
\beq
{\cal L}_{GF} = - {1\over 2\xi} (\partial^M{\Ah}^a_M)^2.
\label{gaugefix}
\eeq
We expect this theory to have high-energy behavior similar to
that of an effective 4D KK gauge theory.
For instance, consider the elastic gauge-boson scattering, 
$
\Ah^a_{j_1}\Ah^b_{j_2} \to \Ah^c_{j_3}\Ah^d_{j_4}
$, 
where $j_k \in (1,2,3)$ denotes the polarization state of the massless
5D gauge field $\Ah_M$.
For an $\mathrm SU(m)$ Yang-Mills theory, we may define
the spin-0, gauge-singlet two-particle state, 
\beqa
|\Psi_0\rangle &=& \dis\f{1}{\sqrt{3}}\f{1}{\sqrt{m^2-1}}
                 \sum^3_{j=1}\sum^{m^2-1}_{a=1}
                 \left|\Ah^a_j \Ah^a_j\right\rangle \,.
\eeqa
The Feynman amplitude for scattering
in this spin-0 gauge-singlet channel takes the form, 
\beqa
{\cal M}_{0}\[\Psi_0 \to \Psi_0\]
&=&
\dis\f{2m}{3}g_5^2\(\f{12}{1-\cos^2\theta}-1\) ,
\eeqa
at the tree-level.  In four dimensions, such a behavior would be unitary to
arbitrarily high energies (so long as $g_5^{~}$ was not too large) reflecting the
renormalizability of 4D Yang-Mills theory. In five dimensions, however, the
properly normalized spin-0 gauge-singlet $s$-partial wave amplitude
\cite{Soldate:1987mk,Chaichian:1988zt} is given by,
\beq
\TT_{00}~=~ {\sqrt{s} \over 64 \pi^2} \int^{\pi}_0 \!d\theta \sin^2\theta\,
{\cal M}_{0} ~=~ \f{23m}{192\pi} \(\,g_5^2\sqrt{s}\,\) \, ,
\label{eq:T005D}
\eeq
where $\sqrt{s}$ is the center-of-mass energy and $\theta$ the center of mass
scattering angle\footnote{Note that, due to the properties of
  five-dimensional phase space, there are no infrared singularities.}.
Unitarity requires that $|{\rm Re} {\TT}_{00}| \le 1/2$, and hence this
amplitude respects tree-level unitarity only for energies
\beq
\sqrt{s} ~=~ E_{cm} ~\le~ {96 \pi \over 23 m}\,{1\over g^2_5}  \,.
\label{eq:Ubound5D}
\eeq
This result is a manifestation that five-dimensional Yang-Mills
theory is, at best, a low-energy effective theory valid only up to
scales of order $1/g^2_5$\,.

We now show that these results can be recovered in compactified
five-dimensional Yang-Mills theory, viewed in four dimensions.  For
convenience, we consider compactifying the fifth-dimension to a line segment
$0 \le x^5 \le \pi R$\,. This can be done consistently by an orbifold projection
as follows: restrict the fields $A^M(x^N)$ to those periodic in $x^5$ with
period $2\pi R$ and further impose a $Z_2$ symmetry,
\beq
\Ah^\mu(x^\nu,x^5) = + \Ah^\mu(x^\nu,-x^5)\,,~~~~ 
\Ah^5(x^\nu,x^5)   = - \Ah^5(x^\nu,-x^5)  \,.
\eeq
These projections force the gauge-covariant 
boundary conditions,
\beq
\widehat{F}^{5N}=\widehat{F}^{N5}=0
\eeq
at $x^5=0$ and $\pi R$.
The theory is then invariant under a restricted
set of gauge transformations
\beq
\Ah^M(x) \to U(x) \Ah^M(x) U^\dagger(x) + 
{i \over g_5^{~}} U(x)\partial^M U^\dagger(x) \,,
\label{gaugeinv}
\eeq
which respect the orbifold projection conditions, i.e. gauge
transformations $U(x) = \exp\[-i g_5^{~} \epsilon^a(x) T^a\]$ for
which $\epsilon^a(x^\mu, x^5) = + \epsilon^a(x^\mu, -x^5)$.

The four-dimensional content of the theory is
most easily seen by expanding $\Ah_{\mu}$ in a Fourier cosine series
\beq
\Ah^a_\mu = {1 \over \sqrt{\pi R}} \left[ A^{a0}_\mu(x_\nu) + 
\sqrt{2}\sum_{n=1}^{\infty} A^{an}_\mu(x_\nu) 
\cos\({n x_5\over R}\)\right]\, ,
\eeq
and $\Ah_5$ in a Fourier sine series
\beq
\Ah^a_5 = \sqrt{2\over \pi R} 
\sum_{n=1}^{\infty} A^{an}_5(x_\nu) \sin\({n x_5\over R}\)\, .
\eeq
In terms of the Fourier expansions, the kinetic energy
terms in the Lagrangian (\ref{fivedym}) become
\beq
{\cal L}_{K.E.} = -{1\over 4} 
\left[\(\partial_{[\mu} A^{a0}_{\nu ]}\)^2
+ \sum_{n=1}^{\infty} \(\partial_{[\mu} A^{an}_{\nu ]}\)^2 \right]
-{1\over 2} \sum_{n=1}^\infty\left[M_n A^{an}_\mu + 
\partial_\mu A^{an}_5\right]^2\,,
\eeq
where \,$M_n=n/R$\, is the mass of the KK state at level-$n$.

The gauge transformations in eqn.\,(\ref{gaugeinv}) allow for the gauge
fixing of the 4D gauge theory [with gauge-bosons $A^{a0}_\mu(x_\mu$)], as well
as allowing us to choose values for the $A^{an}_5(x_\mu)$. We may therefore
impose a general $R_\xi$ gauge-fixing of the form,
\beq
{\cal L}^\prime_{GF} = - \sum_{n=0}^\infty 
{1 \over 2 \xi_n} \left( \partial^\mu A^{an}_\mu 
+ \xi_n M_n A^{an}_5\right)^2~,
\eeq
where the $\{\xi_n\}$ (with $n=0,1,2,\ldots$) are arbitrary gauge parameters.  
 From these expansions, we see that the zero-modes 
$\{A^{a0}_\mu\}$ form an adjoint of
massless vector bosons as expected, while the $\{A^{an}_\mu\}$ form a
Kaluza-Klein (KK) tower of adjoint vector bosons with mass $M_n=n/R$.  The
gauge-fixing term eliminates the kinetic-energy mixing between $A^{an}_\mu$
and $A^{an}_5$, and we may identify the $A^{an}_5$ modes as the ``eaten''
Goldstone bosons of a geometrical ``Higgs'' mechanism where no physical Higgs
boson is actually invoked. The $A_5^{an}$ has a gauge-dependent mass
$M_{5n}^2=\xi_n M_n^2$. The appropriate Faddeev-Popov ghost term 
can be derived, though it is not needed for the analysis below.

The analysis of the compactified theory proceeds most simply in the ``unitary''
gauge, $\xi_n = \infty$ for $n\ge 1$, in which $\{A^{an}_5\}$ fully
decouple since $M_{5n}\to\infty$. The
self-interactions of the zero-mode fields are that of a usual 4D
Yang-Mills theory with gauge-coupling $g = g_5^{~} / \sqrt{\pi R}$
and covariant gauge-fixing parameter $\xi_0$. The
interactions of the KK modes amongst themselves and with the zero mode
gauge-bosons are \cite{Dicus:2000hm,Hill:2000mu},
\beqa
{\cal L}_{\rm int} & = & 
- g C^{abc} \sum_{n=1}^N
\[\partial_{\mu} A^{a0}_{\nu} A^{bn \mu} A^{cn \nu} + 
\partial_{\mu} A^{an}_{\nu} (A^{b0\mu} A^{cn\nu} + A^{bn\mu}A^{c0\nu})\] 
\nonumber 
\\
&& - {g\over \sqrt{2}} C^{abc} \sum_{n,m,l=1}^N \partial_{\mu} A^{an}_{\nu}
A^{bm\mu} A^{c\l\nu} \Delta_3(n,m,\l) 
\nonumber 
\\
&&- {g^2\over 4} C^{abc} C^{ade} \sum_{n=1}^N
\[A^{b0}_\mu A^{c0}_\nu A^{dn\mu} A^{en\nu}
+ {\rm all\ permutations}\] 
\label{selfint}
\\  
&& -{g^2 \over 4\sqrt{2}} C^{abc} C^{ade} \sum_{n,m,\l =1}^N 
\Delta_3(n,m,\l) 
\[A^{b0}_\mu A^{cn}_\nu A^{dm\mu} A^{e\l\nu} + {\rm all\ permutations}\] 
\nonumber 
\\
&& 
-{g^2\over 8} C^{abc} C^{ade} \sum_{n,m,\l,k=1}^N A^{bn}_\mu A^{cm}_\nu
A^{d\l\mu} A^{ek \nu} \Delta_4(n,m,\l,k) ~, \nonumber
%\label{selfint}
\eeqa
with $(\Delta_3,\,\Delta_4)$ given by,
\beqa
\Delta_3 (n,m,\l)
& = & \;\,\delta(n+m-\l)+\delta(n-m+\l)+\delta(n-m-\l) 
\nonumber \\
\Delta_4 (n,m,\l,k)
& = & \;\,\delta(n+m+\l-k)+\delta(n+m-\l+k)+\delta(n-m+\l+k) 
\label{deltas} \\
&& \hspace*{-2mm} 
+\; \delta(n+m-\l-k)+\delta(n-m-\l+k)+\delta(n-m+\l-k)+\delta(n-m-\l-k) \,. ~~~~~
\nonumber
\eeqa
Since the underlying 5D gauge-theory must break down at
energy scale $\cut = \O\(1/g^2_5\)$, we have truncated the KK tower at 
the level $N$ such that $N/R = \Lambda = {\cal O}\(1/g^2_5\)$.

Inspecting eqns.\,(\ref{selfint}) and (\ref{deltas}), 
we see that the KK tower is
a set of self-interacting massive vector bosons, with interactions similar to
those of a four-dimensional massive Yang-Mills theory with a characteristic
coupling $g$. The usual arguments 
\cite{Dicus:1973vj,Lee:1977yc,Lee:1977eg,Veltman:1977rt}
would suggest that the scattering of longitudinally-polarized vector bosons
at level $n$ will grow with energy and would violate unitarity at an energy
scale, 
\beq
\dis
E^\star \sim {4\pi M_n\over g} 
= {4n \pi \over g R} 
= {4n\pi^{3\over2} \over g^{~}_5\sqrt{R}}                              
= {4n\pi^2 g\over g^2_5}~.
\label{eq:UB0}
\eeq
However, this cannot be the case as can be seen in several ways.  First, $g$
could in principle be {\it arbitrarily} small by adjusting $R$, in which case
$E^\star$ could be made {\it arbitrarily} small\footnote{It is interesting to
  note that for $\delta$ extra dimensions the scale $E^\star$ is proportional
  to $g^{2/\delta-1}/g^{2/\delta}_{4+\delta}$, and for six dimensions or
  greater is necessarily smaller than the corresponding $\Lambda$ -- the
  intrinsic scale of the higher dimensional gauge theory -- so long as the
  compactification size is greater than $1/\Lambda$.}. In particular, if this
were the case, $E^\star$ could be made much smaller than the intrinsic cutoff
of the order $1/g^2_5$ (as inferred from our analysis of the 5D Yang-Mills
theory).  Second, the compactification can be viewed as the imposition of the
appropriate boundary conditions on 5D Yang-Mills fields for which, as we have
previously argued, tree-level scattering amplitudes {\it do not} grow with
energy.

In addition, it has recently been shown that the low-energy properties of a
compactified five-dimensional gauge theory may be reproduced in a
``deconstructed'' (or ``remodeled'') four-dimensional effective field theory
with a replicated gauge group and an appropriate gauge-symmetry breaking
pattern \cite{Arkani-Hamed:2001ca,Hill:2000mu}.  These four-dimensional
models may be interpreted as theories in which a compactified fifth dimension
is discretized with a lattice spacing of order $a=R/N$, where $N$ is the
number of replicated gauge groups. Furthermore, these theories can be
embedded in a variety of renormalizable four-dimensional gauge theories
\cite{Arkani-Hamed:2001ca,Hill:2000mu}, in which case it is not possible that
unitarity is violated at any energy.  By making $N$ large (for fixed $a$),
$E^\star$ can be made arbitrarily small.  In particular $E^\star$ can be made
smaller than $1/a$, the energy scale at which the deconstructed theory
deviates significantly from the compactified 5D Yang-Mills theory.  Since the
deconstructed theory cannot violate unitarity at this energy, neither can the
five-dimensional gauge theory\footnote{Scattering in the deconstructed theory
  deviates from compactified 5D Yang-Mills theory by corrections of order $1/N$,
  some of which grow with energy. Vector boson scattering in these theories
  will be explored in a forthcoming publication \cite{Hjhe:2001}.}.

To elucidate this behavior, we consider the elastic scattering of two
longitudinally-polarized KK vector bosons with level-$n$. The relevant
Feynman diagrams are depicted in Fig.\,1, which includes the exchange of the
zero-mode states, the states with level $2n$, as well as the four-point
contact couping of level-$n$ states amongst themselves. A careful analysis of
these contributions\footnote{Details of this and related calculations will be
presented elsewhere \protect\cite{more:2001}.} shows that the individual
terms have energy dependences of $\O(E^4)$ and $\O(E^2)$, but due to
cancellations among all of these diagrams, the overall scattering amplitude
{\it does not} grow with energy. 
Instead, after a lengthy calculation we find that the amplitude approaches a
{\it constant,} 
\beqa
\TT\!\[A^{an}_L A^{bn}_L \!\to\! A^{cn}_L A^{dn}_L\]  \!\!\!&=&\!\!\!
g^2\!\[
C^{abe}C^{cde}\!\(\f{5}{2}c\) +
C^{ace}C^{dbe}\!\(-\f{8c^2-5c+9}{2(1-c)}\) + 
C^{ade}C^{bce}\!\( \f{8c^2+5c+9}{2(1+c)}\) 
\] 
\nonumber\\[3.5mm]
&&
\hspace*{6mm}  +\, \O\!\({M_n^2}/{E^2}\) \,,
\label{hebehavior}
\eeqa
where $c=\ct$.

The cancellations in this amplitude arise from the gauge symmetry of the
underlying five-dimensional theory
(and in particular the Jacobi identity of the structure
constants $C^{abc}$ which ensures the $\O(E^2)$ cancellation), 
as well as the particular masses 
($M_n = n /R$) of the various KK levels.  
The unitarity of this process depends
crucially on the cancellation of contributions from level 
$0$, $n$, and $2n$.
Unlike the traditional Higgs mechanism\,\cite{Higgs:1964ia}, 
in which the unitarity of massive
vector boson scattering is assured through the exchange of a spin-0 Higgs
boson, in the current case 
the unitarity of level-$n$ scattering occurs through the
introduction of a new set (level-$2n$) of {\it vector bosons}! Of course,
unitarity of level-$2n$ scattering would require the addition of higher-level
vector bosons and, ultimately, the {\it entire tower} of KK states.

\begin{figure}
%\vspace*{-12cm}
\begin{center}
\hspace*{-2.3cm}
\includegraphics[width=20.3cm,height=9.3cm]{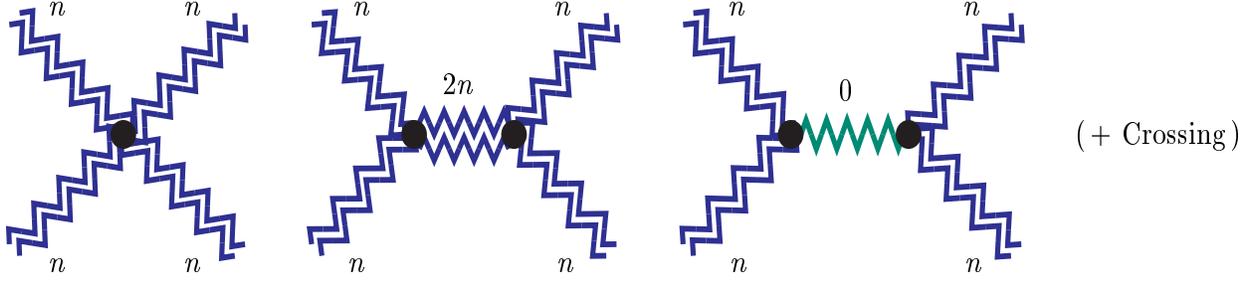} %
\end{center}
\vspace*{-6.3cm}
\caption{Longitudinal KK scattering, 
$A^{an}_LA^{bn}_L \to A^{cn}_LA^{dn}_L$,
in compactified 5D Yang-Mills theory.}
\label{fig:4n}
\end{figure}

The behavior of the high-energy longitudinal KK scattering in the
compactified theory can also be understood from examining the
corresponding Goldstone amplitude of $A^{an}_5$.  We observe that, 
as a consequence of the geometric Higgs mechanism reflected in
the $R_\xi$ gauge-fixing term (\ref{gaugefix}),  the
amplitude of $A_L^{an}$ and that of $A^{an}_5$ are connected via a
Kaluza-Klein Equivalence Theorem (KK-ET) in the high energy limit $E\gg M_n$.
In analogy with the traditional ET in the standard model  
for longitudinal weak gauge boson scattering
\cite{Cornwall:1974km,Lee:1977eg,Chanowitz:1985hj,Yao:1988aj,
      Bagger:1990fc,He:1992ng,He:1994yd,He:1997cm},
we deduce the relation,
\beq
\TT\[A^{an}_L(p_n), A^{bm}_L(p_m), \ldots\]  = C_{\rm mod}\,
\TT\[A^{an}_5(p_n), A^{bm}_5(p_m), \ldots\]  + {\cal O}(M_{n,m,\ldots}/E) \,,
\label{KK-ET}
\eeq
where each external momentum is put on mass-shell, e.g., $p_n^2=M_n^2$, etc,
and the radiative modification factor $C_{\rm mod} = 1+\O({\rm loop})$ arises
only at
loop-level\,\cite{Yao:1988aj,Bagger:1990fc,He:1992ng,He:1994yd,He:1997cm} and
is irrelevant to the tree-level analysis below.  The physical longitudinal
amplitude of $A_L^{an}$ in (\ref{KK-ET}) may be computed in any gauge while
the Goldstone $A^{an}_5$-amplitude only exists in the $R_\xi$ gauges such as
the 't Hooft-Feynman gauge ($\xi_n=1)$ or Landau gauge ($\xi_n=0$).

In the $R_\xi$ gauge, there are additional 
interactions involving the $A^{an}_5$ Goldstone states 
({\it n.b.} $A_5 = -A^5$), which we derive as,
\beqa
{\cal L}^{(5)}_{\rm int} & = & 
+ g C^{abc} \sum_{n=1}^N 
A^{b0\mu} A^{cn}_5 (\partial^\mu A^{an}_5 + M_n A^{an\mu})
+ {g^2 \over 2} C^{abc} C^{ade} \sum_{n=1}^N 
A^{b0}_\mu A^{d0\mu} A^{cn}_5 A^{en}_5 
\nonumber   
\\
&&
+ {g \over \sqrt{2}} C^{abc} \sum_{n,m,\l=1}^N
A^{bn}_\mu A^{cm}_5
(\partial^\mu A^{a\l}_5 + M_{\l} A^{a\l\mu})
\widetilde{\Delta}_3(n,m,\l) 
\nonumber 
\\[-2mm]
&& \label{L5}
\\[-2mm]
&& 
+{g^2\over \sqrt{2}} C^{abc} C^{ade} \sum_{n,m,\l=1}^N 
A^{b0}_\mu A^{dn\mu} A^{cm}_5 A^{e\l}_5 
\widetilde{\Delta}_3(n,m,\l) 
\nonumber 
\\
&& +{g^2\over 4} C^{abc} C^{ade} \sum_{n,m,\l,k=1}^N 
A^{bn}_\mu A^{dm\mu} A^{c\l}_5 A^{ek}_5 
\widetilde{\Delta}_4(n,m,\l,k) \,,
\nonumber 
\eeqa
where
\beqa
\widetilde{\Delta}_3(n,m,\l) \!\!\!& = &\!\!
\,\delta(n+m-\l) + \delta(n-m+\l) - \delta(n-m-\l)\,,
\nonumber 
\\
\widetilde{\Delta}_4(n,m,\l,k) \!\!\!& = &\!\! 
\,\delta(n+m+\l-k) +\delta(n+m-\l+k)
+\delta(n-m+\l-k) + \delta(n-m-\l+k)~~~~~~~  
\\
\!\!\!&&\!\!
\hspace*{-2mm}
-\delta(n+m-\l-k) - \delta(n-m+\l+k) - \delta(n-m-\l-k) 
\nonumber \,. 
\eeqa
 From this equation we see that the states $A^{an}_5$ interact as a set of
color-octet scalar particles, and their cubic (quartic) vertices contain only
one (zero) partial derivative and one or two (two) Goldstone fields of
$A^{an}_5$. Power-counting therefore shows that the $A^{an}_5$ amplitude is
at most of $\O(E^0)$ and is manifestly unitary in four-dimensions.  Based
upon the equivalence theorem (\ref{KK-ET}), this should reproduce the same
high-energy behavior (\ref{hebehavior}) for longitudinal KK scattering. Using
(\ref{L5}) we explicitly compute the $A^{an}_5$ amplitude to be
\beqa
\TT\[A^{an}_5 A^{an}_5 \to  A^{an}_5 A^{an}_5 \]
\!\!\!&=&\!\!\!
g^2\!\[
C^{abe}C^{cde}\!\(-\f{3}{2}c\) +
C^{ace}C^{dbe}\!\(-\f{3(3+c)}{2(1-c)}\) + 
C^{ade}C^{bce}\!\( \f{3(3-c)}{2(1+c)}\) 
\] 
\nonumber\\[3.5mm]
&&
\hspace*{6mm}  +\, \O\!\({M_n^2}/{E^2}\)~.
\label{Amp-A5}
\eeqa
This differs from (\ref{hebehavior}) only by an overall constant
$-4c$ times the Jacobi identity, 
\beq
C^{abe}C^{cde} + C^{ace}C^{dbe} + C^{ade}C^{bce} = 0 \,,
\label{Jacobi}
\eeq
and thus perfectly agrees with the KK-ET in eqn.\,(\ref{KK-ET}).

While it is reassuring that the low-energy unitarity of elastic scattering in
the five-dimensional Yang-Mills theory is reproduced in the four-dimensional
compactified theory, it is natural to wonder how the {\it bad} high-energy
behavior of the underlying five-dimensional theory is manifest in the
compactified theory.  In fact, the bad high-energy behavior of the underlying
theory is manifest not in the behavior of a single scattering channel, but
rather in a coupled-channel analysis.  Consider energies
large compared to the mass of the level-$N_0$ KK modes, and the (normalized)
state consisting of equal parts of pairs of longitudinally polarized gauge
bosons from all of the first $N_0$ levels,
\beq
\left|\Psi^{ab}\right\rangle = 
{1 \over \sqrt{N_0}} \sum_{\l=1}^{N_0} 
\left| A^{a\l}_L A^{b\l}_L \right\rangle \,.
\eeq
We then compute the inelastic amplitude,
\beqa
\TT\!\[A^{an}_L A^{bn}_L \to A^{cm}_L A^{dm}_L\]\!
\!\!\!&=&\!\!\!
g^2\!\[
C^{abe}C^{cde}\!\(-c\) +
C^{ace}C^{dbe}\!\(-\f{3+c}{1-c}\) + 
C^{ade}C^{bce}\!\( \f{3-c}{1+c}\)\!
\]  
+ \O\!\( \!\f{M_{n,m}^2}{E^2}\!\! \)\!  
\nonumber\\
&=&  \dis\TT\!\[A^{an}_5 A^{bn}_5 \to A^{cm}_5 A^{dm}_5\]
     +\O\( {M_{n,m}^2}/{E^2} \) 
\label{nn-mm} \\
&=& \dis\f{2}{3} \TT\!\[A^{an}_5 A^{bn}_5 \to A^{cn}_5 A^{dn}_5\]
   +\O\( {M_{n,m}^2}/{E^2} \) \,,
\nonumber
%\label{nn-mm}
\eeqa
where we have verified again 
that the longitudinal and Goldstone amplitudes are
equivalent in the high energy limit and 
differ only by terms of $\O({M_{n,m}^2}/{E^2})$.
 From these,  we arrive at
\beqa
\TT \[ \left|\Psi^{ab}\right\rangle 
\to \left|\Psi^{cd}\right\rangle \] 
& \simeq & (N_0-1)
  \TT\[A^{an}_L A^{bn}_L \to A^{c\l}_L A^{d\l}_L\]
+ \TT\[A^{an}_L A^{bn}_L \to A^{cn}_L A^{dn}_L\] 
\nonumber\\[-2mm]
\\[-2mm]
&\simeq & N_0 \TT\[A^{an}_5 A^{bn}_5 \to A^{c\l}_5 A^{d\l}_5\]         
         +\O\({M_{n,\l}^2}/{E^2}\) 
,~~~~   ({\rm for}~~N_0 \gg 1)\,.
\nonumber
\eeqa
So, for large $N_0$, the normalized four-dimensional gauge-singlet 
$s$-wave amplitude is,
\beqa
\TT_\Psi^{00} &=&\dis \f{1}{64\pi}\int_{-1}^1 \!d\ct\,
\f{1}{m^2-1}\sum^{m^2-1}_{a,c=1}
\TT \[ \left|\Psi^{aa}\right\rangle \to 
       \left|\Psi^{cc}\right\rangle \] \nonumber \\
&\simeq&
\dis N_0\f{mg^2}{16\pi}\[-1+2\ln\f{s}{|M_n^2-M_\ell^2|}\]
= N_0{ mg^2 \over 16 \pi}\O(1) 
= \f{N_0}{R}{mg^2_5 \over 16 \pi^2}\O(1) ,
\label{eq:T00N0}
\eeqa
where we have retained the pole masses in the $(t,u)$-channel contributions
to the inelastic Goldstone amplitude  
in order to avoid the infrared singularity in the phase space. 
The associated logarithmic factor is of $\O(1)$.
Requiring the $s$-wave amplitude in eqn.\,(\ref{eq:T00N0}) to
be less than $1/2$, we find that the KK tower must be
truncated at the level $N_0=N$ such that
\beq
{ N \over R} ~\,\lesssim~\, {8 \pi^2 \over m}\f{1}{g_5^2} ~.
\label{eq:UboundKK}
\eeq
As in our discussion of unitarity of the 5D Yang-Mills theory
[cf. eqn.\,(\ref{eq:Ubound5D})], 
we see that the compactified 4D KK theory must be
treated as an effective theory valid only below a scale of the order
$1/g^2_5$.  Unlike our expectation based on massive 4D Yang-Mills
theory [cf. eqn.\,(\ref{eq:UB0})], 
the bound has no dependence on the effective 4D gauge
coupling $g\,(=g^{~}_5/\sqrt{\pi R}\,)$ 
and is therefore independent of the radius of compactification.

%%%%%%%%%%%%%%%%%%%%%%%%%%%%%%
\vspace*{4mm}
\noindent
{\bf Acknowledgments}\\[2mm]
We are grateful to B.~Dobrescu, J.~Distler, H.~Georgi, and S.~Nandi for
discussions.  This work was supported in part by the Department of Energy
under grants DE-FG02-91ER40676 and DE-FG03-93ER40757.

%%%%%%%%%%%%%%%%%%%%%%%%%%%%%%%
\vspace*{4mm}
\noindent
{\bf Note added:}~~ As this work was being completed, we became aware a new
preprint\,\protect\cite{new:2001} which also considered $R_\xi$ gauge-fixing
and the resulting Feynman rules in compactified Yang-Mills theory.

%%%%%%%%%%%%%%%%%%%%%%%%%%%%%%%%%%%%%%%%%%%%%%%%%%%%%%%%%%%%%%%%%%%%%

\bibliography{unitary.bib}

\begin{thebibliography}{10}

\bibitem{Antoniadis:1990ew}
I.~Antoniadis,
\newblock Phys. Lett. {\bf B246}, 377 (1990).
%%CITATION = PHLTA,B246,377;%%

\bibitem{Antoniadis:1993fh}
I.~Antoniadis, C.~Munoz, and M.~Quiros,
\newblock Nucl. Phys. {\bf B397}, 515 (1993), hep-ph/9211309.
%%CITATION = HEP-PH 9211309;%%

\bibitem{Lykken:1996fj}
J.~D. Lykken,
\newblock Phys. Rev. {\bf D54}, 3693 (1996), hep-th/9603133.
%%CITATION = HEP-TH 9603133;%%

\bibitem{Antoniadis:1997hk}
I.~Antoniadis and M.~Quiros,
\newblock Phys. Lett. {\bf B392}, 61 (1997), hep-th/9609209.
%%CITATION = HEP-TH 9609209;%%

\bibitem{Arkani-Hamed:1998rs}
N.~Arkani-Hamed, S.~Dimopoulos, and G.~R. Dvali,
\newblock Phys. Lett. {\bf B429}, 263 (1998), hep-ph/9803315.
%%CITATION = HEP-PH 9803315;%%

\bibitem{Antoniadis:1998ig}
I.~Antoniadis, N.~Arkani-Hamed, S.~Dimopoulos, and G.~R. Dvali,
\newblock Phys. Lett. {\bf B436}, 257 (1998), hep-ph/9804398.
%%CITATION = HEP-PH 9804398;%%

\bibitem{Shiu:1998pa}
G.~Shiu and S.~H.~H. Tye,
\newblock Phys. Rev. {\bf D58}, 106007 (1998), hep-th/9805157.
%%CITATION = HEP-TH 9805157;%%

\bibitem{Dienes:1998vh}
K.~R. Dienes, E.~Dudas, and T.~Gherghetta,
\newblock Phys. Lett. {\bf B436}, 55 (1998), hep-ph/9803466.
%%CITATION = HEP-PH 9803466;%%

\bibitem{Dienes:1998vg}
K.~R. Dienes, E.~Dudas, and T.~Gherghetta,
\newblock Nucl. Phys. {\bf B537}, 47 (1999), hep-ph/9806292.
%%CITATION = HEP-PH 9806292;%%

\bibitem{Pomarol:1998sd}
A.~Pomarol and M.~Quiros,
\newblock Phys. Lett. {\bf B438}, 255 (1998), hep-ph/9806263.
%%CITATION = HEP-PH 9806263;%%

\bibitem{Cheng:1999fu}
H.-C. Cheng, B.~A. Dobrescu, and C.~T. Hill,
\newblock Nucl. Phys. {\bf B573}, 597 (2000), hep-ph/9906327.
%%CITATION = HEP-PH 9906327;%%

\bibitem{Lykken:1999ms}
J.~Lykken and S.~Nandi,
\newblock Phys. Lett. {\bf B485}, 224 (2000), hep-ph/9908505.
%%CITATION = HEP-PH 9908505;%%

\bibitem{LlewellynSmith:1973ey}
C.~H. Llewellyn~Smith,
\newblock Phys. Lett. {\bf B46}, 233 (1973).
%%CITATION = PHLTA,B46,233;%%

\bibitem{Dicus:1973vj}
D.~A. Dicus and V.~S. Mathur,
\newblock Phys. Rev. {\bf D7}, 3111 (1973).
%%CITATION = PHRVA,D7,3111;%%

\bibitem{Cornwall:1973tb}
J.~M. Cornwall, D.~N. Levin, and G.~Tiktopoulos,
\newblock Phys. Rev. Lett. {\bf 30}, 1268 (1973).
%%CITATION = PRLTA,30,1268;%%

\bibitem{Cornwall:1974km}
J.~M. Cornwall, D.~N. Levin, and G.~Tiktopoulos,
\newblock Phys. Rev. {\bf D10}, 1145 (1974).
%%CITATION = PHRVA,D10,1145;%%

\bibitem{Lee:1977yc}
B.~W. Lee, C.~Quigg, and H.~B. Thacker,
\newblock Phys. Rev. Lett. {\bf 38}, 883 (1977).
%%CITATION = PRLTA,38,883;%%

\bibitem{Lee:1977eg}
B.~W. Lee, C.~Quigg, and H.~B. Thacker,
\newblock Phys. Rev. {\bf D16}, 1519 (1977).
%%CITATION = PHRVA,D16,1519;%%

\bibitem{Veltman:1977rt}
M.~J.~G. Veltman,
\newblock Acta Phys. Polon. {\bf B8}, 475 (1977).
%%CITATION = APPOA,B8,475;%%

\bibitem{Soldate:1987mk}
M.~Soldate,
\newblock Phys. Lett. {\bf B186}, 321 (1987).
%%CITATION = PHLTA,B186,321;%%

\bibitem{Chaichian:1988zt}
M.~Chaichian and J.~Fischer,
\newblock Nucl. Phys. {\bf B303}, 557 (1988).
%%CITATION = NUPHA,B303,557;%%

\bibitem{Dicus:2000hm}
D.~A. Dicus, C.~D. McMullen, and S.~Nandi,
\newblock (2000), hep-ph/0012259.
%%CITATION = HEP-PH 0012259;%%

\bibitem{Hill:2000mu}
C.~T. Hill, S.~Pokorski, and J.~Wang,
\newblock Phys. Rev. {\bf D64}, 105005 (2001), hep-th/0104035.
%%CITATION = HEP-TH 0104035;%%

\bibitem{Arkani-Hamed:2001ca}
N.~Arkani-Hamed, A.~G. Cohen, and H.~Georgi,
\newblock Phys. Rev. Lett. {\bf 86}, 4757 (2001), hep-th/0104005.
%%CITATION = HEP-TH 0104005;%%

\bibitem{Hjhe:2001}
R.~S. Chivukula and H.-J. He,
\newblock (2001),
\newblock to be published.

\bibitem{more:2001}
R.~S. Chivukula, D.~A. Dicus, and H.-J. He,
\newblock (2001),
\newblock work in preparation.

\bibitem{Higgs:1964ia}
P.~W. Higgs,
\newblock Phys. Lett. {\bf 12}, 132 (1964).
%%CITATION = PHLTA,12,132;%%

\bibitem{Chanowitz:1985hj}
M.~S. Chanowitz and M.~K. Gaillard,
\newblock Nucl. Phys. {\bf B261}, 379 (1985).
%%CITATION = NUPHA,B261,379;%%

\bibitem{Yao:1988aj}
Y.-P. Yao and C.-P. Yuan,
\newblock Phys. Rev. {\bf D38}, 2237 (1988).
%%CITATION = PHRVA,D38,2237;%%

\bibitem{Bagger:1990fc}
J.~Bagger and C.~Schmidt,
\newblock Phys. Rev. {\bf D41}, 264 (1990).
%%CITATION = PHRVA,D41,264;%%

\bibitem{He:1992ng}
H.-J. He, Y.-P. Kuang, and X.~Li,
\newblock Phys. Rev. Lett. {\bf 69}, 2619 (1992).
%%CITATION = PRLTA,69,2619;%%

\bibitem{He:1994yd}
H.-J. He, Y.-P. Kuang, and X.~Li,
\newblock Phys. Rev. {\bf D49}, 4842 (1994).
%%CITATION = PHRVA,D49,4842;%%

\bibitem{He:1997cm}
H.-J. He and W.~B. Kilgore,
\newblock Phys. Rev. {\bf D55}, 1515 (1997), hep-ph/9609326.
%%CITATION = HEP-PH 9609326;%%

\bibitem{new:2001}
A.~Muck, A.~Pilaftsis, and R.~Ruckl,
\newblock (2001), hep-ph/0110391.
%%CITATION = HEP-PH 0110391;%%

\end{thebibliography}
\bibliographystyle{h-physrev3}

\end{document}